\documentstyle[epsfig,aps,floats,preprint]{revtex}

\newcommand{\beq}{\begin{equation}}
\newcommand{\eeq}{\end{equation}}
\newcommand{\bea}{\begin{eqnarray}}
\newcommand{\eea}{\end{eqnarray}}

\def\laq{~\raise 0.4ex\hbox{$<$}\kern -0.8em\lower 0.62
ex\hbox{$\sim$}~}
\def\gaq{~\raise 0.4ex\hbox{$>$}\kern -0.7em\lower 0.62
ex\hbox{$\sim$}~}

\def \ra {\rightarrow}

\def \a {\alpha}

\def \ga {\gamma}

\def \ep {\epsilon}

\def \om {\omega}
\def \Om {\Omega}

\begin{document}
\par
\begingroup

\begin{flushright}
BA-TH/00-398\\
gr-qc/0101118\\
\end{flushright}
\vskip 1true cm

\vspace{20mm}
{\large\bf\centering\ignorespaces
  A diagrammatic approach to the spectrum \\ of cosmological
perturbations
\vskip2.5pt}

\bigskip
{\dimen0=-\prevdepth \advance\dimen0 by23pt
\nointerlineskip \rm\centering
\vrule height\dimen0 width0pt\relax\ignorespaces
G. De Risi and M. Gasperini
\par}

{\small\it\centering\ignorespaces
Dipartimento di Fisica, Universit\`a di Bari, \\
Via G. Amendola 173, 70126 Bari, Italy \\
and \\Istituto Nazionale di Fisica Nucleare, Sezione di Bari,
Bari, Italy \\
\par}

\par
\bgroup
\leftskip=0.10753\textwidth \rightskip\leftskip
\dimen0=-\prevdepth \advance\dimen0 by17.5pt \nointerlineskip
\small\vrule width 0pt height\dimen0 \relax

\begin{abstract}
We compute the spectral distribution of the quantum fluctuations of the 
vacuum, amplified by inflation, after an arbitrary number of background 
transitions. Using a graphic representation of the process we find that
the  final spectrum can be completely determined trough a synthetic
set of working  rules, and a list of simple algebraic computations.
\end{abstract}

\vspace{5mm}
\begin{center}
---------------------------------------------\\
\vspace {5 mm}

To appear in {\bf Phys. Lett. B}
\end{center}


\par\egroup
\thispagestyle{plain}
\endgroup



The standard cosmological model \cite{1}, in spite of the large number
of successful  predictions, cannot be extrapolated too far back in time.
It is by now a consolitated opinion  that our Universe, in its earlier
epochs, should include a phase of inflationary (i.e.  accelerated)
evolution  \cite{2}. Many different realizations of inflation are
possible,  however, and we have to face the problem of how to
distinguish them through their possible  phenomenological
consequences.

To this aim, we may note that the instability of the small fluctuations
(of the metric and of  the matter fields) is one of the main physical
properties of a phase of accelerated  evolution. As a consequence of this
instability, small perturbations are amplified in a way  which depends on
the intensity and on the duration of inflation \cite{3}. Different
inflationary  models amplify perturbations with different spectral
distribution, and the computation of the  spectrum thus becomes an
important tool to predict observable effects, and to discriminate 
eventually between various possible models of primordial evolution.

The computation of the spectrum, performed accorting to the
standard cosmological  perturbation theory \cite{3}, may become long
and cumbersome even in the linear  approximation, however, if the
model of cosmological background is complicated. The aim of  this paper
is to present a set of working rules, based on a diagrammatic
representation of  the evolution of perturbations, allowing a quick
estimate of the final amplitude and of the spectral  distribution. The
idea of a diagrammatic approach is not new \cite{4}, but it was never 
implemented up to now in a complete and systematic way, so as to be
generally valid for all  classes of inflationary backgrounds.

To be more precise, the diagrammatic method of computation that we
shall propose here can be  applied to all cosmological fluctuations
$\psi$ described by the quadratic effective action 
\beq
S={1\over 2} \int d\eta~z^2(\eta) \left( \psi^{\prime 2} + \psi 
\nabla^2\psi \right), 
\label{1}
\eeq
where $z$ is the external ``pump field", responsible for the
amplification, and the prime  denotes differentiation with respect to the
conformal time $\eta$. It is convenient, in this  context, to introduce the
canonical variable $u=z\psi$ which diagonalizes the effective  action,
and which satisfies the canonical evolution equation \cite{3} 
\beq
u^{''}_{k}+\left(k^2-{z^{\prime\prime}\over z}\right)u_k=0
\label{2}
\eeq
(for each mode $k$ of the Fourier expansion, $\nabla^2u_k=-k^2u_k$). 
We shall also assume that the  background evolution can be separated
into $n+1$ different cosmological phases, with $n$ transitions  at
$\eta=\eta_i$, $i= 1,...,n$, and that in each phase the evolution of the
pump fields (sufficiently far  from the transition) can be
parametrized in conformal time by an appropriate power $\a_i$, namely 
$z_i\sim|\eta|^{\a_i}$. It follows that, in each phase, the general
solution of eq. (\ref{2}) can be  written in terms of the first  and second
kind Hankel functions \cite{5} as: 
\beq  
u^i_k(\eta)=|\eta|^{1 \over 2}
\left[A^i_+(k)H^{(1)}_{\mu_i}(x)+A^i_-(k)H^{(2)}_{\mu_i}(x)\right], 
~~~~~\mu_i=|\a_i-{1 /2}|,~~~~~x= k|\eta|.
\label{3}
\eeq

In order to fix our conventions, we shall label the $n$ transitions in
decreasing order for $\eta$  ranging from $-\infty$ to $+\infty$, and we
shall put the power $\a_i$ to the right (and $\a_{i+1}$  to the left) of
$\eta_i$, namely 
\beq 
z_{i+1}\sim|\eta|^{\a_{i+1}},~~~~\eta<\eta_i, ~~~~~~~~~~~~~~
z_{i}\sim|\eta|^{\a_{i}},~~~~\eta>\eta_i
\label{4}
\eeq
It follows that the first (in order of time) cosmological phase is labelled
by $n+1$, the last one by $1$. The normalization to an initial vacuum
fluctuation spectrum \cite{3} thus imposes $A_+^{n+1}=1$, 
$A_-^{n+1}=0$, while the coefficients of the last phase $A^1_\pm(k)$,
fixed by the continuity of  $\psi_k$ and $\psi'_k$ at the transitions 
$\eta_i$ ({\em not} of $u_k$ and $u'_k$, see the discussion  in \cite{6}),
will determine the final spectral distribution of the amplified
perturbations. The  spectral energy density, in particular, is given by
\cite{3} $d\rho/d\log k=(k/a)^4|A^1_-(k)|^2/\pi^2$ or, in critical units
and in terms of the proper frequency $\om=k/a$,  \beq
\Om(\om,t)={8 \over 3\pi} {\om^4 \over M^2_pH^2 } |A^1_-(\om)|^2.
\label{4a} 
\eeq 

The computation of the spectrum thus reduces, in
general, to the problem of solving a linear  non-omogeneous system of
$2n$ equations, 
\beq
\psi_k^{i+1}(x_i)=\psi_k^{i}(x_i),~~~~~~~~~~~~
\psi_k^{\prime ~i+1}(x_i)=\psi_k^{\prime ~i}(x_i), 
\label{5}
\eeq
for the $2n$ unknown quantities $A^i_\pm(k)$.  The solution is
straightforward, in principle; in practice,  however, it is in general
arduous to extract the relevant physical information from the exact
solution,   for any background with $n\geq2$. In this paper we shall
derive a set of prescriptions enabling an  immediate estimate of the
spectral coefficient $A^1_-(k)$, through an approximate procedure
which captures  the essential features (amplitude and frequency
dependence) of the spectrum, without solving  explicitly the full system
of equations. Such an estimate is based on the asymptotic expansion of
the  Hankel functions, and on the possible separation of the spectrum
into $n$ frequency bands, depending on the  number of transitions which
are truly effective for a given frequency mode.

In order to introduce the basic ideas of our procedure, we shall start
considering the simplest case of  only one transition at $\eta=\eta_1$.
The normalized solution of eq. (\ref{2}) is then 
\bea
&&
u^1_k=|\eta|^{{1 \over 2}}H^{(1)}_{\mu_1}(x), \nonumber \\
&&
u^2_k=|\eta|^{{1 \over 2}}\left[A^1_+H^{(1)}_{\mu_2}(x)
+A^1_-H^{(2)}_{\mu_2}(x) \right] 
\label{6}
\eea
(with $\mu_1\neq\mu_2$), and the continuity of $\psi_k$ at $\eta_1$
provides for the spectral coefficients  the following exact solution
\cite{6}: 
\bea
&&
A^1_-={i\pi \over 4}\left[x_1
\left(H^{'(1)}_{\mu_2}H^{(1)}_{\mu_1}-H^{(1)}_{\mu_2}H^{'(1)}_{\mu_1}
\right)  + \left(\a_1-\a_2\right)H^{(1)}_{\mu_2}H^{(1)}_{\mu_1}\right],
\nonumber \\ 
&& A^1_+={i\pi \over 4}\left[x_1
\left(H^{(1)}_{\mu_2}H^{'(2)}_{\mu_1}-H^{'(1)}_{\mu_2}
H^{(2)}_{\mu_1}\right)  +
\left(\a_2-\a_1\right)H^{(1)}_{\mu_2}H^{(2)}_{\mu_1}\right], 
\label{7}
\eea  
where the Hankel functions are evaluated ay $x_1=k|\eta_1|$. These 
coefficients satisfy the canonical  normalization $|A_+|^2-|A_-|^2=1$. 

The time scale $\eta_1$ defines the typical transition frequency,
$k_1=|\eta_1|^{-1}$. For modes with  $k \gg k_1$ we can use the large
argument limit of the Hankel functions, and we obtain $A^1_-\simeq0$, 
$A^1_+\simeq1$: such modes are thus unaffected by the transition,
modulo higher order corrections that are  esponentially damped like
$e^{-k/k_1}$ (see \cite{7}, for instance), and that we shall neglect in
our  approximation. For low frequency modes, $k\ll k_1$, we can use
instead the small argument limit \cite{5}, 
\bea &&
H^{(1)}_{\mu}(x)\simeq 
p_{\mu}x^{\mu}+iq_{\mu}x^{-\mu}-ir_{\mu}x^{2-\mu}+
s_{\mu}x^{2+\mu}+..., ~~~~~~~~~x\rightarrow0 \nonumber \\ 
&&
H^{(2)}_{\mu}(x)\simeq 
p_{\mu}^*x^{\mu}-iq_{\mu}x^{-\mu}+ir_{\mu}x^{2-\mu}+
s_{\mu}^*x^{2+\mu}+...,  ~~~~~~~~~x\rightarrow0
\label{8} 
\eea
where the coefficients $p,q,...$ are complex numbers with  modulo of
order one (for later use, we have  also included higher order
corrections). The exact solution (\ref{7}) provides, in this limit, 
\beq
A^1_-\simeq 
C^1_1x_1^{-\mu_2-\mu_1}+C^1_2x_1^{-\mu_2+\mu_1}
+C^1_3x_1^{\mu_2-\mu_1}+C^1_4x_1^{2-\mu_2-\mu_1}+... 
\label{9}
\eeq
where
\bea
&
C^1_1={i\pi \over 4} q_{\mu_1}q_{\mu_2}\left(\mu_2-\mu_1+
\a_2-\a_1\right), \nonumber \\
&
C^1_2=-{\pi \over 4} p_{\mu_1}q_{\mu_2}\left(-\mu_2-\mu_1-
\a_2+\a_1\right), \nonumber \\
&
C^1_3=-{\pi \over 4} q_{\mu_1}p_{\mu_2}\left(\mu_2+\mu_1-
\a_2+\a_1\right), \nonumber \\
&
C^1_4={i\pi \over 4} \left[q_{\mu_1}r_{\mu_2}\left(2-\mu_2+
\mu_1-\a_2+\a_1\right) + 
r_{\mu_1}q_{\mu_2}\left(-2-\mu_2+\mu_1-\a_2+\a_1\right)\right]
\label{10}
\eea
(and a similar expression for $A^1_-$). 

By recalling that $\mu_i=|\a_i-1/2|$, it follows that the first term of
the expansion (\ref{9}) is  the leading one, for $x_1\rightarrow0$. When
$C^1_1=0$, however, we have to include the  next-to-leading
corrections. Taking into account all possible values of $\a_1$, $\a_2$,
and  truncating the expansion (\ref{9}) to the lowest order term with
non-vanishing coefficients, we find  that for $\a_1\neq\a_2$ there are
four different possibilities, corresponding to four different  spectral
amplitudes: 
\bea &
\a_2>1/2~{\rm or}~\a_1>1/2,~~~~~~~~~~
&
A^1_- \simeq C^1_{1} x_1^{-\mu_2}x_1^{-\mu_1}, \nonumber \\
&
\a_1>-1/2,~\a_1>\a_2,~~~~~~~~~~
&
A^1_- \simeq C^1_{2}x_1^{-\mu_2}x_1^{\mu_1}, \nonumber \\
&
\a_2>-1/2,~\a_2>\a_1,~~~~~~~~~~
&
A^1_- \simeq C^1_{3} x_1^{\mu_2}x_1^{-\mu_1}, \nonumber \\
&
\a_2\leq-1/2,\a_1\leq-1/2,~~~~~~~~~~
&
A^1_- \simeq C^1_{4} x_1^{-\mu_2+1}x_1^{-\mu_1+1}, 
\label{11}
\eea
in agreement with the results first obtained in \cite{6}.

If one of the two powers $\a_1$, $\a_2$ equals $1/2$, 
the corresponding Bessel index is $\mu=0$, and 
the small argument expansion (\ref{8}) is to be replaced by
\bea
&&
H^{(1)}_{0}(x) \simeq 
p_{0}+iq_{0}\log x-ir_{0}x^{2}\log x+s_{0}x^{2}+..., 
~~~~~~~x \ra 0,\nonumber \\
&&
H^{(2)}_{0}(x)\simeq 
p_{0}^{*}-iq_{0}\log x+ir_{0}x^{2}\log x+s_{0}^{*}x^{2}+...,
~~~~~~~x \ra 0.
\label{12}
\eea
Suppose, for instance, that $\a_1=1/2$. The exact solution (\ref{7}) is
 now approximated by
\beq
A^1_-\simeq 
C^1_1x_1^{-\mu_2}\log x_1 +\overline{C}^1_2 x_1^{-\mu_2}+... ,
\label{13}
\eeq
where
\beq
\overline{C}^1_2=- \frac{\pi} {4} \left[q_{\mu_2}p_0 \left( -\mu_2 -
\a_2 + {1 \over 2} \right) -iq_{\mu_2}  q_0 \right].
\label{14}
\eeq
As $\overline{C}^1_1$ is always nonzero, the leading term
 in eq. (\ref{13}) is the firs one for $\a_2>1/2$, 
and the second one for $\a_2<1/2$. We may thus include also 
the value $\a=1/2$ in the general rules 
(\ref{11}), provided we take into account the prescription
\beq
x^{-\mu}|_{\mu=0}\rightarrow \log x,~~~~~~~~ x^{\mu}|_{\mu=0}
\rightarrow 1.
\label{sost}
\eeq

The above computation  for a single background
transition can be easily iterated for a  cosmological model containing
two or more transitions. We have solved the general case with $n$ 
transitions, and we have found the remarkable result that the vanishing
of the leading term of the  asymptotic expansion - and then the
particular spectral behaviour of the solution - depends only on  the
kinematic powers $\a_i$ of two phases: the one \em{preceding the
first} transition, and the one  \em{following the last} transition. Such a
result is in agreement with the well known phenomenon of  ``freezing"
of perturbations \cite{3}, and with the general duality properties
\cite{8} of the action  (\ref{1}). 

Using the above result, it becomes possible to write down a recurrent
expansion for the spectrum after  $n$ transition. To this aim, it is
convenient to represent the whole amplification process with a  simple
diagram, in which we insert a vertical line in correspondence of each
transition. The height  of the {\em i-th} line is proportional to the
associated transition frequency, $k_i=|\eta_i|^{-1}$. It  become
possible, in this way, to identify at a glance the various frequency
bands ($k$) of the  spectrum, according to the number of transitions
(with $k_i>k$) from which a given band is  significatively affected. Note
that, for $\eta$ growing from minus infinity, the height of the  vertical
lines may grow  monothonically up to a maximum transition frequency
(corresponding to a  minimum time scale), and is then monothonically
decreasing for $\eta$ running towards plus infinity.

Let us start with the case in which the transition frequencies 
$k_i=|\eta_i|^{-1}$ are arranged in 
growing order  from the left to the right (see Fig. 1).
The frequency band with $k>k_1$ is not amplified and we shall
disregard it. The  band number $1$ of the 
diagram ($k_2<k<k_1$) will be affected by one transition, the band
number $2$  ($k_3<k<k_2$) by two transitions, and so on. The iteration
of the matching procedure used for a  single transition leads 
to a recurrent expression for the coefficient $A_-$, where the
$k$-dependence  is fixed by the first 
and last phase, and the amplitude is fixed by the continuity at the 
transitions. We can write, in particular,
\bea
|A^1_- (k)| & \simeq 
& 
x_1^{\ga_{21}}x_1^{\ga_{12}},
~~~~~~~~~~~~~~~~~~~~~~~~~~~~~~~~~~~~~~~~~~~~~~~~~~
~~~~~~~~~~ k_2<k<k_1   \nonumber \\
& 
\simeq & x_2^{\ga_{31}}x_1^{\ga_{13}} 
\left(\frac{x_1}{x_2}\right)^{\ga_{21}}\left(\frac{x_1}{x_2}
\right)^{\ga_{12}-\ga_{13}},~~~~~~~~~~~~~~~~~~~~~~~~~~~~~~~~~
k_3<k<k_2  \nonumber \\ 
& 
\vdots & \nonumber \\
& 
\simeq & x_n^{\ga_{n+1,1}}x_1^{\ga_{1,n+1}} 
\left(\frac{x_{n-1}}{x_n}\right)^{\ga_{n1}}\left(\frac{x_1}{x_n}
\right)^{\ga_{1
n}-\ga_{1,n+1}} ... \nonumber \\
&&
... 
\left(\frac{x_{i-1}}{x_i}\right)^{\ga_{i1}}\left(\frac{x_1}{x_i}
\right)^{\ga_{1i}-\ga_{1,i+1}} ... 
\left(\frac{x_1}{x_2}\right)^{\ga_{21}}\left(\frac{x_1}{x_2}
\right)^{\ga_{12} - 
\ga_{13}},~~~~~~  k<k_n 
\label{15}
\eea
(modulo a numerical factor of order one, that can be computed from
the exact solution, but that we shall neglect for our purpose of a quick
approximate estimate). 

\begin{figure}[t]
\begin{center}
\mbox{\epsfig{file=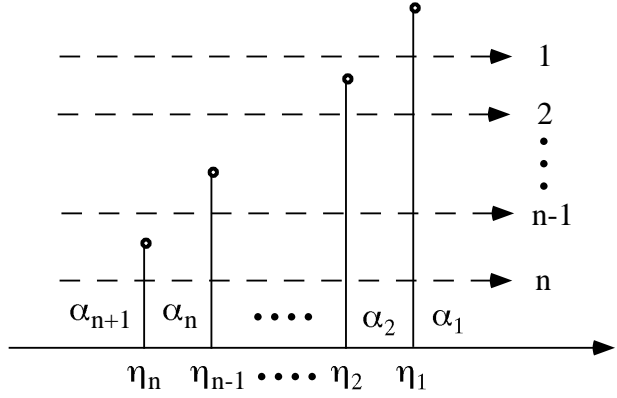,width=82mm}}
\vskip 5mm
\caption{\sl The $n$ frequency bands of the spectrum for a background
with $n$ transition frequencies, arranged in growing order.}
\end{center}
\end{figure}

The powers $\ga_{ik}$ of eq. (\ref{15}) depend (in ordered way)
on $\a_i,\a_k$ according to the same rules of eq. (\ref{11}), with the
only difference that when $\a_i$ and $\a_k$ are not contiguous
(i.e., $\a_i\not=\a_{k+1}$), they can also have the same value. In that
case $\mu_i=\mu_k$, and the lowest order non-vanishing term of the
expansion leads to $x_{i-1}^{-\mu_i}x_k^{\mu_i}$ if $x_{i-1}<x_k$, 
or to  $x_{i-1}^{\mu_i}x_k^{-\mu_i}$ if $x_{i-1}>x_k$ (see eq. (\ref{10}),
where $C_2^1=-C_3^1 \not=0$ for $\mu_1 =\mu_2$). By summarizing
the working rules for computing the powers $\ga_{ik}$ we must then
distinguish two cases. In the case $\a_i=\a_k$ we have: 
\bea
\a_i=\a_k , ~~~\Rightarrow ~~~ &&\ga_{ik} =\ep_{ik} \mu_i, ~~~~~~~~~
\ga_{ki} =-\ep_{ik} \mu_i,
\nonumber\\
&&
\ep_{ik} = {\rm sign}\log \left(x_{i-1}\over x_k\right)=
{\rm sign} \log \left(k_{k}\over k_{i-1}\right). 
\label{16}
\eea
Otherwise ($\a_i \not= \a_k$) we have, according to eq. (\ref{11}): 
\bea &
\a_i>1/2~{\rm or}~\a_k>1/2, ~~~~\Rightarrow ~~~~
&
\ga_{ik}=-\mu_i, ~~~~~~~~~~~~~\ga_{ki}=-\mu_k, \nonumber \\
&
\a_i>-1/2,~\a_i>\a_k, ~~~~~~\Rightarrow ~~~~
&
\ga_{ik}=\mu_i, ~~~~~~~~~~~~~~~\ga_{ki}=-\mu_k, \nonumber \\
&
\a_k>-1/2,~\a_k>\a_i, ~~~~~~\Rightarrow ~~~~
&
\ga_{ik}=-\mu_i, ~~~~~~~~~~~~\ga_{ki}=\mu_k, \nonumber \\
&
\a_i\leq-1/2,\a_k\leq-1/2, ~~~~\Rightarrow ~~~~
&
\ga_{ik}=-\mu_i+1, ~~~~~~~\ga_{ki}=-\mu_k+1, \nonumber \\
\label{17}
\eea

Note that $\ga_{ik}\not= \ga_{ki}$, but that $\ga_{ik}(\a_i,\a_k)= 
\ga_{ik}(\a_k,\a_i)$. Note also that, in the limiting case in which
$\mu_i=0$ or $\mu_k=0$, we can take into account the logarithmic
corrections according to the prescription (\ref{sost}). However, as 
they are usually negligible in realistic models, we shall neglect the log
corrections in our first estimate of the spectrum, using the simple
rule:
\beq
\ga_{ik}\left({1\over 2},\a_k\right) =0, ~~~~~~~~~~~~~~~~
\ga_{ki}\left({1\over 2},\a_k\right) =-\mu_k=\left |\a_k-{1\over
2}\right|,  
\label{18}
\eeq
for any $k$. 

\begin{figure}[t]
\begin{center}
\mbox{\epsfig{file=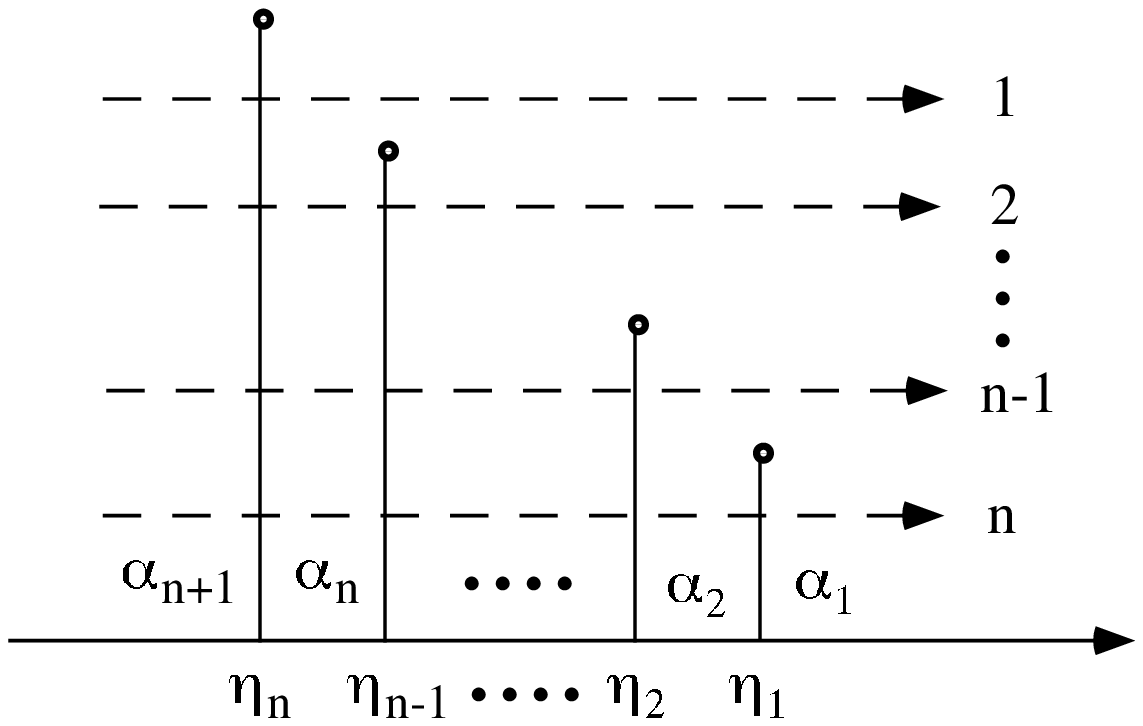,width=82mm}}
\vskip 5mm
\caption{\sl The $n$ frequency bands of the spectrum for a background
with $n$ transitions frequencies, arranged in decreasing order.}
\end{center}
\end{figure}

At this point, one remark is in order. The spectral
distribution determined through the above procedure applies -- by
construction -- only sufficiently far from the transition frequencies, $k
\ll k_i$. Near the transition the spectral slope may change with respect
to the asymptotic regime, but always in such a way as to guarantee the
continuity of the spectrum at $k=k_i$. In eq. (\ref{15}), however, the
asymptotic expression has been extrapolated up to $k_i$ and, as a
consequence, its amplitude is normalized by continuity. This is certainly
allowed  -- within our approximations -- when the leading term of the
expansion (\ref{9}) is nonzero. When $\a_i$ and $\a_1$ are both
smaller than $1/2$, and the leading term is vanishing, this
extrapolation may introduce an error in the asymptotic amplitude of the
spectrum, which is however of order one -- and thus compatible with
the degree of accuracy required  for our present estimate -- provided
the amplification of the corresponding band is not too large, i.e.
provided $F(k_i) \equiv  (k_i/k_{i-1})^{\ga_{i1}}(k_i/k_{1})^{\ga_{1i}}
\laq 1$. If $F \gg1$, on the contrary, the correct asymptotic amplitude
of the branches with $k <k_i$ is obtained by renormalizing the results of
eq. (\ref{15}) by the factor $F^{-2}(k_i)$.

A recurrent expression, similar to eq. (\ref{15}), is also obtained for the
complementary situation in which the $n$ transition frequencies $k_i$
are arranged in decreasing order, for $\eta$ ranging from minus to plus
infinity (see Fig. 2). There are still $n$ frequency bands, defined by
$k_{i-1}<k<k_i$,  $i=1, ... , n$, and the spectral coefficients can be
approximated as follows: 
\bea
|A^1_- (k)| & \simeq 
& 
x_n^{\ga_{n+1,n}}x_n^{\ga_{n,n+1}},
~~~~~~~~~~~~~~~~~~~~~~~~~~~~~~~~~~~~~~~~~~~~~~~~~~
~~~~~~~~~~~~ k_{n-1}<k<k_n  \nonumber \\
& 
\simeq &x_n^{\ga_{n+1,n-1}}x_{n-1}^{\ga_{n-1,n+1}}
\left(\frac{x_n}{x_{n-1}}\right)^
{\ga_{n,n+1}}\left(\frac{x_n}{x_{n-1}}
\right)^{\ga_{n+1,n}-\ga_{n+1,n-1} }, ~~~~~~~~~~
k_{n-2}<k<k_{n-1}  \nonumber \\ 
& 
\vdots & \nonumber \\
& 
\simeq & x_n^{\ga_{n+1,1}}x_1^{\ga_{1,n+1}} 
\left(\frac{x_{n}}{x_{n-1}}\right)^{\ga_{n,n+1}}\left(\frac{x_n}{x_{n-1}}
\right)^{\ga_{n+1,n}-\ga_{n+1,n-1}}  \nonumber \\
&& 
\left(\frac{x_{n-1}}{x_{n-2}}\right)^{\ga_{n-1,n+1}}\left(\frac{x_n}
{x_{n-2}} \right)^{\ga_{n+1,n-1}-\ga_{n+1,n-2}} ... 
\left(\frac{x_2}{x_1}\right)^{\ga_{2,n+1}}\left(\frac{x_n}{x_1}
\right)^{\ga_{n+1,2} - \ga_{n+1,1}}, \nonumber \\
&&
~~~~~~~~~~~~~~~~~~~~~~~~~~~~~~~~~~~~~~~~~~~~~~
~~~~~~~~~~~~~~~~~~~~~~~~~~~~~~~~~~ k<k_1 
\label{19}
\eea
where $\ga_{ik}$ are given again by eqs. (\ref{16}), (\ref{17}). Again,
when the leading term is vanishing, the asymptotic amplitudes are
correct provided ${\overline F}(k_i) \equiv 
(k_i/k_{i+1})^{\ga_{i+1,n+1}}(k_i/k_{n})^{\ga_{n+1,i+1}} \laq 1$,
otherwise they are to be renormalized by the factor ${\overline
F}^{-2}(k_i)$.

The above results can be summarized by a set of prescriptions,
allowing an automatic computation of the spectrum, once the relevant
diagram is plotted. For a synthetic formulation of such prescriptions it
is convenient to define, for each ``leg" of the diagram representing a
transition, the so-called phase of exit and phase of re-enter. More
precisely, for any leg $\eta_i$ placed to the left of the highest one, we
shall define {\em phase of re-enter} the one {\em directly to the right}
of the {\em last leg} crossed by the transition frequency
$k_i=|\eta_\i|^{-1}$. For any leg $\eta_j$ placed to the right of the
highest one, we shall define {\em phase of exit} the one {\em
directly to the left} of the {\em first leg} crossed by the transition
frequency $k_j=|\eta_j|^{-1}$ (see Fig. 3). 

With the above definitions, the diagrammatic rules for the computation
of the spectrum can now be synthetized as follows. 

\begin{itemize}

\item{} We plot the diagram for the given model of background

\item{}We choose the frequency band we want to compute, and we
single out the relevant transitions. 

\item{}For any relevant leg $\eta_i$ placed to the left of the highest
one we insert the amplitude factor:
\beq
\left(x_{i-1}\over x_i\right)^{\ga_{ir}}
\left(x_{r}\over x_i\right)^{\ga_{ri}-\ga_{r,i+1}}.
\label{20}
\eeq
For any relevant leg $\eta_i$ placed to the right of the highest
one we insert the  factor:
\beq
\left(x_{i+1}\over x_i\right)^{\ga_{i+1,e}}
\left(x_{e-1}\over x_i\right)^{\ga_{e,i+1}-\ga_{ei}}, 
\label{21}
\eeq
where the labels $``e"$ and $``r"$ denote, respectively, the exit and
re-enter phase (note that, according to these rules, the highest leg
does not contribute to the amplitude). 

\item{} We add the overal factor determing the frequency dependence
of the spectrum:
\beq
x_f^{\ga_{f+1,\ell}}x_\ell^{\ga_{\ell,f+1}},
\label{22}
\eeq
where the labels $``f"$ and $``\ell"$ denote, respectively, the {\em
first}  and {\em last} (from the left) relevant legs for the band we are
considering. 

\item{} We compute, finally, the powers $\ga_{ik}$ according to eqs.
(\ref{16}), (\ref{17}) and, if needed, we renormalize the amplitudes
through the factors $F^{-2}$ or ${\overline F}^{-2}$, as discussed
before. 
\end{itemize}

\begin{figure}[t]
\begin{center}
\mbox{\epsfig{file=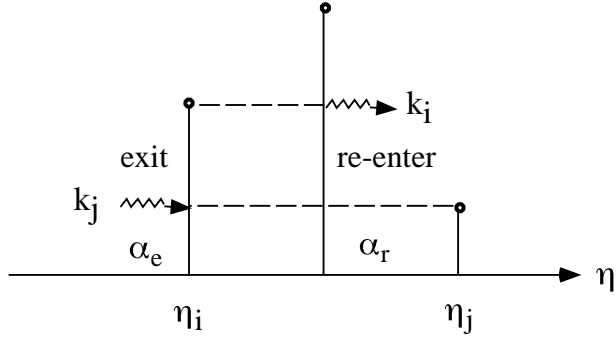,width=82mm}}
\vskip 5mm
\caption{\sl A graphic representation of the exit phase for the ``leg"
$\eta_j$, and of the re-enter phase for the ``leg"
$\eta_i$.}
\end{center}
\end{figure}

As a simple application of such a method, we shall compute here the
spectrum for the amplification process represented by the diagram of
Fig. 4, by assuming for the various phases the following particular
values: 
\beq
\a_1=1, ~~~~~~~~\a_2=1/2, ~~~~~~~~\a_3=2, ~~~~~~~~\a_4=1/2.
\label{22a}
\eeq
With such a choice of the kinematical powers, the diagram of Fig. 4
may represent (in the Einstein frame) the amplification of tensor
metric fluctuations in a non-minimal model of string cosmology
\cite{4,9}, which includes an initial dilaton-driven phase
($\eta<\eta_3$), a first intermediate, high-curvature string  phase
($\eta_3<\eta<\eta_2$), a second intermediate dilaton phase with
decreasing curvature ($\eta_2<\eta<\eta_1$), and a final
radiation-dominated phase, with constant dilaton ($\eta>\eta_1$). In
the limit $\eta_2 \ra \eta_1$ the model reduces to the minimal one, in
the limit $\eta_3 \ra \eta_2$ one recovers instead the non-minimal
model discussed in \cite{6}. 

\begin{figure}[t]
\begin{center}
\mbox{\epsfig{file=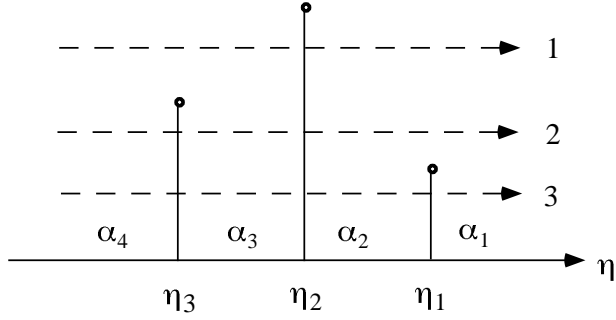,width=82mm}}
\vskip 5mm
\caption{\sl An example of diagram representing the amplification of
tensor metric perturbations, in a non-minimal model of string
cosmology.}
\end{center}
\end{figure}

Let us consider, for instance, the (lowest frequency) band number $3$ of
Fig. 4, corresponding to a ``three-leg" transition (for this band,
obviously, $\eta_3$ is the first leg and $\eta_1$ is the last one). For the
leg  $\eta_3$ the re-enter phase is $\a_2$, for the leg $\eta_1$ the exit
phase is $\a_4$. Following the rules listed above we obtain the spectral
coefficient
\beq
|A_-^1(k)| \simeq x_3^{\ga_{41}}x_1^{\ga_{14}}
\left(x_2\over x_3\right)^{\ga_{32}}\left(x_2\over
x_3\right)^{\ga_{23}-\ga_{24}} \left(x_2\over x_1\right)^{\ga_{24}}
\left(x_3\over x_1\right)^{\ga_{42}-\ga_{41}}, ~~~~~~ k<k_3.
\label{23}
\eeq
The computation of the powers $\ga_{ik}$, according to eqs. (\ref{16}), 
(\ref{17}), (\ref{18}), gives:
\bea 
&
\ga_{32}\left(2,{1\over 2}\right)&=-{3\over 2}, ~~~~~~~~~
\ga_{23}\left(2,{1\over 2}\right)=0, \nonumber \\
&
\ga_{24}\left({1\over 2},{1\over 2}\right)&=0, ~~~~~~~~~~~~
\ga_{42}\left({1\over 2},{1\over 2}\right)=0, \nonumber \\
&
\ga_{41}\left({1\over 2},1\right)&=0, ~~~~~~~~~~~~
\ga_{14}\left({1\over 2},1\right)=-{1\over 2}, 
\label{24}
\eea
where the ordered labels of $\ga$ refers to the various phases, and the
numbers enclosed in round brackets refer to the particular numerical
values of the corresponding kinematical powers. The spectral
coefficient (\ref{23}) thus becomes
\beq
|A_-^1(k)| \simeq \left(k_3\over k_2\right)^{-3/2}
\left(k\over k_1\right)^{-1/2},
\label{25}
\eeq
and the associated energy distribution (\ref{4a}), 
\beq
\Om(\om,t)\simeq {8\over 3 \pi} {\om_1^4\over M_p^2 H^2}
\left(\om_2\over \om_3\right)^{3}
\left(\om\over \om_1\right)^{3} , 
\label{26}
\eeq
reproduces (modulo logarithmic corrections) the well known cubic
slope associated to the dilaton-radiation transiton, and is in agreement
with the results of \cite{4} and \cite{9}. With the same procedure we
can easily estimate the spectrum for the other bands of Fig. 4. 

In summary, we have shown in this paper how to obtain a quick
estimate of the cosmological spectra using a simple method, based on a
set of effective prescriptions, and on a diagrammatic representation
of the amplification of perturbations. We believe that such a method
does not  represent only a mathematical ``curiosity", but may have
useful applications to the study of  various inflationary scenarios,
that will be discussed in future papers. 

\acknowledgments
It is a pleasure to thank  Alessandra Buonanno, Massimo
Giovannini, Carlo Ungarelli and Gabriele Veneziano for helpful comments.

\end{document}